\title{Analysis of Large Scale Web Experiments Using Sequences of Estimators}
\author{
        Ian E. Fellows (Streamlet Data)
}
\date{\today}

\documentclass{article}
\usepackage{dsfont}
\usepackage{natbib}
\usepackage{amsmath}
\usepackage{MnSymbol}

\usepackage[margin=1.5in]{geometry}

\newtheorem{theorem}{Theorem}
\newtheorem{definition}{Definition}
\newtheorem{corollary}{Corollary}

\begin{document}
\maketitle

\begin{abstract}
Experimental testing is vital in the optimization of web applications, and as such A/B testing has been widely adopted as a methodology for determining optimal content for many web applications. While some testing platforms provide sequentially valid inferences, a large proportion of online tests still utilize traditional statistical tests that do not allow for interim ``peeking'' at the data or extending the test past its proposed sample size.

In this paper we develop results useful for the sequential analysis of large scale experiments. In particular, the properties of sequences of maximum likelihood and generalized method of moments estimators are examined. This leads to new tests of odds ratios and relative risks for binary outcomes. For continuous and ordinal outcome we develop a test of mean difference and a non-parametric test of Area Under the Curve (AUC). Additionally, multivariate versions of these tests are proposed.

\end{abstract}

\section{Introduction}

Experimental testing is vital in the optimization of web applications, and as such A/B testing has been widely adopted as a methodology for determining optimal content for many web applications. A/B testing, from a statistical perspective, is a randomized controlled trial (RCT), where users are randomized to one of many user experiences. The goal of these trials is to drive users into particular behaviors such as signing up to a service (a conversion), or purchase of a product.

Despite the widespread adoption of A/B testing, the methodology used to make decisions often does not fit the analyst's needs or actions. Traditional hypothesis tests require the collection of a sample with fixed size, with no interim analysis and no post-hoc study continuation. However, in the context of application optimization, data streams in over the course of days or weeks and discussion making needs to be flexible and dynamic. For this reason, sequentially valid hypothesis tests are required, where significance can be evaluated at any point during the run of the experiment, and type I error is controlled regardless of the number of interim analyses and length of the experiment.

In Section 2 we review the mixture sequential likelihood ratio (mSPRT) and define asymptotic equivalence. Section 3 provides results examining the asymptotic behavior of mSPRT type ratios for sequences of maximum likelihood or generalized method of moment estimators (GMM). Section 4 proposed an efficient family of mixture distributions that can be calculated in closed form. Section 5 applies the theoretical results of Section 4 to the risk ratio, odds ratio and difference of proportions metrics for binary outcomes and the difference of means and AUC metrics for non-binary outcomes.

\section{Preliminaries: Sequential Likelihood Ratio Tests}

Suppose that we wish to test the hypothesis $H_0 : \theta_1 = \theta_0$ for some family of probability distributions $f(x\mid\theta)$, where $\theta$ is a vector of parameters. In a sequential experiment, we observe a sequence of observations from this distribution $X_1,X_2,...X_\infty$, and wish to determine a stopping rule $T$, which may at any point in the sequence reject the null hypothesis and terminate the experiment. The sequential likelihood ratio of $\theta_1$ versus $\theta_0$ is defined as 
$$
L_n = \prod_{i=1}^n \frac{f(X_1,...,X_n\mid\theta_1)} {f(X_1,...,X_n\mid\theta_0) }
$$

Analogous to the likelihood ratio test in classical statistics, \cite{wald1945} suggested rejecting $H_0$ when the sequential likelihood ratio rose above a certain level. Based on the fact that $L_n$ is a martingale, \cite{wald1945} derived type I error probabilities for this test based on the identity
$$
P_{\theta_0}(\textrm{max} \ L_n \ge \alpha^{-1}) \le \alpha.
$$ 
This identity guarantees that rejecting the null hypothesis when the likelihood ratio attains a value of $1/\alpha$ provides an error rate less than or equal to $\alpha$.

For the majority of real world analyses, the hypotheses are composite rather than simple, which complicates the problem considerably. Researchers have considered a number of different generalizations to the Wald sequential likelihood ratio, including the adaptive likelihood ratio \citep{robbins1970a} and the sequential generalized likelihood ratio \citep{schwarz1962} (see \cite{lai2004} for a review). For this work we consider the mixture likelihood ratio test (mSPRT) \citep{robbins1970}, which averages the numerator over a specified prior distribution ($g$) for $\theta$
$$
\Lambda_n^g = \frac{\int \prod_i^nf(X_i\mid\theta)g(\theta)d\theta}{\prod_i^nf(X_i\mid\theta_0)}.
$$
The mixture likelihood ratio then rejects the null hypothesis when
$$
T(\alpha) = \Lambda_n^g > \frac{1}{\alpha},
$$
and terminates the experiment and rejects the null hypothesis at sample size
$$
\tau_G = \textrm{inf}\bigg\{n \ge 1 : \Lambda_n^g > \frac{1}{\alpha}  \bigg\}.
$$

Like $L_n$, $\Lambda_n^g$ is a martingale, and thus the mixture likelihood ratio test is guaranteed to be a level alpha test, in that under the null hypothesis the probability of terminating is 
$$
P(\tau_G < \infty) \le \alpha.
$$
Note that the inequality here is generally fairly tight as the ``overshoot'' of sequential tests is typically small \citep{lai2004}.

For composite null hypothesis, \cite{wald1945} suggested integrating over the denominator as well as the numerator, yielding a likelihood ratio of the form
$$
\Psi_n^{g_1g_2} = \frac{\int \prod_i^nf(X_i\mid\theta)g_2(\theta)d\theta}{\int \prod_i^nf(X_i\mid\theta)g_1(\theta)d\theta},
$$
where $g_2$ is the distribution of the parameters under the alternate hypothesis and $g_1$ is the distribution under the null hypothesis. Rejecting the null when $\textrm{max} \ \Psi^{g_1g_2}_n \ge \alpha^{-1}$ limits the averaged type one error over the null distribution $\int \alpha(\theta)g_1(\theta)d\theta < \alpha$, where $\alpha(\theta)$ is the type I error for the particular paramter configuration $\theta$ \citep{lai2004}.

For large scale experiments we are particularly interested in the large sample behavior of the mSPRT and asymptotic approximations of it. Examination of the large sample characteristics of sequential tests has a long history, starting with \cite{bartlett1946} and \cite{cox1963} with early work summarized by \cite{joanes1972}. We now introduce two definitions relevant for large sample approximations to the mSPRT

\begin{definition} \label{def:valid}
Let $\Lambda_n^g$ be the mSPRT testing the null hypothesis $H_0 : \theta  =  \theta^0$. A sequence $\Lambda_n'$ is asymptotically valid for the mSPRT $\Lambda_n$ if
$$
\Lambda_n' \overset{d}{\Rightarrow} \Lambda_n^g
$$
under $H_0$.
\end{definition}

\begin{definition} \label{def:equiv}
Let $\Lambda_n^g$ be the mSPRT testing the null hypothesis $H_0 : \theta  =  \theta^0$. A sequence $\Lambda_n'$ is asymptotically equivalent to the mSPRT $\Lambda_n$ if
$$
\Lambda_n' \overset{d}{\Rightarrow} \Lambda_n^g
$$
under any value of $\theta$.
\end{definition}

Tests which are asymptotically equivalent to an mSPRT perform similarly at large sample sizes to the underlying mSPRT, while asymptotically valid approximations only perform similarly under the null hypothesis. Obviously, asymptotically equivalent test statistics are also asymptotically valid. 

Tests which are asymptotically valid do not necessarily keep the same type I error characteristics of the mSPRT. At low sample sizes, the asymptotic test may reject too much or not enough. There are two approaches that alleviate this behavior. Firstly, it may be prudent to ignore rejections at low sample sizes and only trigger a termination of the test once a threshold sample size has been reached. Secondly, if $g$ is heavily concentrated around $\theta^0$ then rejections at low sample size are unlikely even under an approximation of $\Lambda^g_n$, and thus inflated type I errors are of little practical concern.

\section{Sequences of Estimators}

Let $\hat{\theta}_n(X_1,...,X_n)$ be a consistent estimator for the parameter vector $\theta$, which conditional upon $X_1,...,X_{n-1}$ is a one-to-one function of $X_n$. Given a known sampling distribution for $\hat{\theta}_n$, it is desirable to create a valid sequential hypothesis test based on this distribution.

\begin{theorem} \label{the:est}
Let $H_0 : \theta = \theta_0$ be a statistical hypothesis and $\hat{\theta}_n$ be sufficient for $\theta$, then the mSPRT is equal to
$$
\Lambda_n^g = \frac{ \int p(\hat{\theta}_n \mid \theta)  g(\theta)d\theta} { p(\hat{\theta}_n \mid \theta_0) }.
$$
\end{theorem}
\textbf{Proof:}
Let $S_n = n\hat{\theta}_n - (n-1)\hat{\theta}_{n-1}$, then
\begin{align*}
\prod_i^nf(X_i\mid\theta) &=f(X_1,...,X_n \mid\theta) \\
&= p(S_1,...,S_n \mid \theta) \\
&= p(\sum^n_iS_i \mid \theta)p(S_1, ..., S_n \mid \sum^n_iS_i, \theta) \\
&=p(\hat{\theta}_n \mid \theta)p(S_1, ..., S_n \mid \hat{\theta}_n).
\end{align*}
Because the estimator is sufficient for $\theta$, the second probability cancels out in the likelihood ratio, leaving
\begin{align*}
\Lambda_n^g &= \frac{\int \prod_i^nf(X_i\mid\theta)g(\theta)d\theta}{\prod_i^nf(X_i\mid\theta_0)} \\
&=\frac{ \int p(\hat{\theta}_n \mid \theta)  g(\theta)d\theta} { p(\hat{\theta}_n \mid \theta_0) }.
\end{align*}
$\blacksquare$

Theorem \ref{the:est} allows for a significant reduction in computational complexity when it can be applied, as there are often easy closed form approximations of the distribution of $\hat{\theta}$. 

\begin{theorem} \label{the:asym}
Let $H_0 : \theta = \theta_0$ be a statistical hypothesis, and $\hat{\theta}_n$ be the maximum likelihood estimators for $\theta$. Further, let $\sqrt{n}\hat{\Sigma}_n^{-\frac{1}{2}}(\hat{\theta}_n-\theta) \overset{d}{\Rightarrow} \mathcal{N}(0, I)$, where $I$ is the identity matrix and $\hat{\Sigma}_n$ is a consistent estimate of the limiting covariance when $H_0$ is true ( $\Sigma(\theta^0)$ ).
$$
\frac{ \int \phi(\hat{\theta}_n \mid \theta , n^{-1}\hat{\Sigma}_n )  g(\theta ) d\theta} { \phi(\hat{\theta}_n \mid \theta^0, n^{-1}\hat{\Sigma}_n ) }
$$
is asymptotically valid for $\Lambda_n^g$, and is asymptotically equivalent if $\hat{\Sigma}_n$ is a consistent for all $\theta$.
\end{theorem}
\textbf{Proof:}

Let $\hat{\theta}_n$ represent the maximum likelihood estimators for $f$, then the Taylor expansion around the MLE is
$$
\log f(\theta) = \log f(\hat{\theta}) + (\theta-\hat{\theta}_n)^T\nabla f \big|_{\hat{\theta}_n}  + \frac{1}{2}(\theta-\hat{\theta}_n)^T\nabla^2 f \big|_{\hat{\theta}_n}(\theta-\hat{\theta}_n) + ...,
$$
with higher order terms tending to 0 as $n$ increases. Using the fact that $\nabla f \big|_{\hat{\theta}_n}=0$, we express the log likelihood ratio as
\begin{align*}
\log f(\theta) - \log f(\theta^0) &= \log f(\theta) - \log f(\hat{\theta}_n)  - (\log f(\theta^0) - \log f(\hat{\theta}_n)) \\
&\Rightarrow -\frac{1}{2}(\theta-\hat{\theta}_n)^T\nabla^2 f \big|_{\hat{\theta}_n}(\theta-\hat{\theta}_n) + \frac{1}{2}(\theta^0-\hat{\theta}_n)^T\nabla^2 f \big|_{\hat{\theta}_n}(\theta^0-\hat{\theta}_n).
\end{align*}
Exponentiating both sides yields
\begin{align*}
\int \frac{f(X_1,...,X_n | \theta)}{f(X_1,...,X_n | \theta^0)} g(\theta)d\theta &\Rightarrow \frac{ \int \phi(\hat{\theta}_n \mid \theta , -(\nabla^2f \big|_{\hat{\theta}_n})^{-1} )  g(\theta ) d\theta} { \phi(\hat{\theta}_n \mid \theta^0, -(\nabla^2f \big|_{\hat{\theta}_n})^{-1} ) }
\end{align*}
If $\hat{\Sigma}_n \Rightarrow -n(\nabla^2f \big|_{\hat{\theta}_n})^{-1}$ equivalence is achieved by substituting $n^{-1}\hat{\Sigma}_n$ into the ratio. If it is only consistent under the null hypothesis then the ratio is valid for $\Lambda_n^g$.

%Let $S_n = n\hat{\theta}_{n} - (n-1)\hat{\theta}_{n-1}$ and consider the ratio
%$$
%T_n=\frac{ \int q(S_1,...,S_n | \theta)  g(\theta) d\theta} { p(S_1,...,S_n | \theta^0) },
%$$
%where $q(S_1,...,S_n | \theta) = p(S_1,...,S_n | \sum_i^n S_i, \theta)\phi(\sum_i^n S_i | \theta, n^{-1}\Sigma(\theta_0))$. In order for $T_n$ to be a valid mSPRT, it must be a martingale. We can see that this is the case as
%\begin{align*}
%E(T_n | S_1,...,S_{n-1}) &= \int \int \frac{  q(S_1,...,S_n | \theta)} { p(S_1,...,S_n | \theta^0) } p(S_n | S_1,...,S_{n-1}, \theta^0)  g(\theta)dS_n d\theta \\
%&= \int \frac{  q(S_1,...,S_{n-1} | \theta)} { p(S_1,...,S_{n-1} | \theta^0) }\Bigg(\int q(S_n | S_1,..., S_{n-1}) dS_n \Bigg)  g(\theta) d\theta \\
%&= T_{n-1}.
%\end{align*}
%We now must show that, under the null hypothesis, as $n$ becomes large, our desired test statistics converges to $T_n$. Asymptotically, $\hat{\theta}$ is sufficient, and thus $p(S_1,...,S_n | \sum_i^n S_i, \theta) = p(S_1,...,S_n | \sum_i^n S_i)$. We then have that
%\begin{align*}
%T_n &\Rightarrow \frac{ \int p(S_1,...,S_n | \sum_i^n S_i, )\phi(\sum_i^n S_i | \theta, n^{-1}\Sigma(\theta_0))  g(\theta) d\theta} { p(S_1,...,S_n | \sum_i^n S_i)p(\sum_i^n S_i | \theta^0) }\\
%&\Rightarrow \frac{ \int \phi(\sum_i^n S_i | \theta, n^{-1}\Sigma(\theta_0))  g(\theta) d\theta} { \phi(\sum_i^n S_i | \theta^0, n^{-1}\Sigma(\theta_0)) }.
%\end{align*}
%The result then follows from the fact that $\hat{\Sigma}_n \Rightarrow \Sigma(\theta^0)$ under the null hypothesis.
$\blacksquare$

\begin{theorem} \label{the:comp}
Let $\theta=\begin{bmatrix} \beta \\ \eta \end{bmatrix}$ be a partitioning of the parameter space, $H_0 : \beta  =  \beta^0$ be the hypothesis of interest and
$$
\Lambda_n = \frac{ \int \phi(\hat{\theta}_n \mid \theta , n^{-1}\hat{\Sigma}_n )  g(\beta) d\beta} { \phi(\hat{\theta}_n \mid \theta^0, n^{-1}\hat{\Sigma}_n ) }
$$
be an asymptotically valid mSPRT, where $\hat{\Sigma}_n = \begin{bmatrix} \hat{\Sigma}^{\beta \beta}_n & \hat{\Sigma}^{\beta \eta}_n \\ \hat{\Sigma}^{\eta \beta}_n & \hat{\Sigma}^{\eta \eta}_n \end{bmatrix}$ is a consistent estimate of the limiting covariance matrix and $\theta^0=\begin{bmatrix} \beta^0 \\ \eta \end{bmatrix}$. Further, let $\hat{\Sigma}^{\eta \theta}=\hat{\Sigma}^{\theta \eta}=0$. 
Then 
$$
\Lambda_n = \frac{ \int \phi(\hat{\beta}_n \mid \beta , n^{-1}\hat{\Sigma}_n^{\beta \beta} )  g(\beta) d\theta} { \phi(\hat{\beta}_n \mid \beta^0, n^{-1}\hat{\Sigma}_n^{\beta \beta} ) }
$$
\end{theorem}
\textbf{Proof:}

\begin{align*}
\frac{ \int \phi(\hat{\theta}_n \mid \theta , n^{-1}\hat{\Sigma}_n )  g(\beta) d\beta} { \phi(\hat{\theta}_n \mid \theta^0, n^{-1}\hat{\Sigma}_n ) } &= \frac{ \int \phi(\hat{\eta}_n \mid \eta , n^{-1}\hat{\Sigma}_n^{\eta \eta} ) \phi(\hat{\beta}_n \mid \beta , n^{-1}\hat{\Sigma}_n^{\beta \beta} )  g(\beta) d\theta} { \phi(\hat{\eta}_n \mid \eta , n^{-1}\hat{\Sigma}_n^{\eta \eta} ) \phi(\hat{\beta}_n \mid \beta^0, n^{-1}\hat{\Sigma}_n^{\beta \beta} ) } \\
&= \frac{ \int \phi(\hat{\beta}_n \mid \beta , n^{-1}\hat{\Sigma}_n^{\beta \beta} )  g(\beta) d\beta} { \phi(\hat{\beta}_n \mid \beta^0, n^{-1}\hat{\Sigma}_n^{\beta \beta} ) }
\end{align*}

$\blacksquare$

In many cases, the assumption that $\hat{\Sigma}^{\theta \eta}=0$ is not met, making inference in the case of nuisance parameters difficult. Let $h$ be the true prior distribution of $\eta$. Averaging the null hypothesis over $h$ yields a likelihood ratio of
$$
\Psi_n^{h(gh)} = \frac{\int \int \prod_i^nf(X_i\mid \beta, \eta)g(\beta)h(\eta)d\eta d\beta}{\int \prod_i^nf(X_i\mid \beta_0, \eta)h(\eta)d\eta}.
$$
Using this ratio as the test statistic limits type I error averaged across all experiments to the specified level. Of course $h$ is unknown, making direct use of $\Psi$ problematic. Fortunately, the following theorem shows that asymptotically, the ratio does not depend on $h$.

\begin{theorem} \label{the:comp2}
Let $\theta=\begin{bmatrix} \beta \\ \eta \end{bmatrix}$ be a partitioning of the parameter space, $H_0 : \beta  =  \beta^0$ be the hypothesis of interest and $\hat{\theta}_n$ be the maximum likelihood estimators for $\theta$. Further, let $\sqrt{n}\hat{\Sigma}_n^{-\frac{1}{2}}(\hat{\theta}_n-\theta) \overset{d}{\Rightarrow} \mathcal{N}(0, I)$, where $I$ is the identity matrix and $\hat{\Sigma}_n = \begin{bmatrix} \hat{\Sigma}^{\beta \beta}_n & \hat{\Sigma}^{\beta \eta}_n \\ \hat{\Sigma}^{\eta \beta}_n & \hat{\Sigma}^{\eta \eta}_n \end{bmatrix}$ is a consistent estimate of the limiting covariance $\Sigma(\theta)$.

(1) If $h(\eta) \propto 1$, then
$$
\frac{ \int \phi(\hat{\beta}_n \mid \beta , n^{-1}\hat{\Sigma}_n^{\beta \beta} )  g(\beta) d\theta} { \phi(\hat{\beta}_n \mid \beta^0, n^{-1}\hat{\Sigma}_n^{\beta \beta} ) }
$$
is asymptotically equivalent to $\Psi_n^{h(gh)}$.

(2) If $h'(\eta | \beta) = h(\eta + \Sigma^{\eta^0 \beta^0}(\Sigma^{\beta^0\beta^0})^{-1} (\beta_0 -\beta))$ then
$$
\frac{ \int \phi(\hat{\beta}_n \mid \beta , n^{-1}\hat{\Sigma}_n^{\beta \beta} )  g(\beta) d\theta} { \phi(\hat{\beta}_n \mid \beta^0, n^{-1}\hat{\Sigma}_n^{\beta \beta} ) }
$$
is asymptotically valid for $\Psi_n^{h(gh')}$ at $\theta^0=\begin{bmatrix} \beta^0 \\ \eta^0 \end{bmatrix}$.

\end{theorem}
\textbf{Proof:}
As with Theorem \ref{the:asym}, we use the fact that
$$
\log f(\theta) - \log f(\hat{\theta}) \Rightarrow  \frac{1}{2}(\theta-\hat{\theta}_n)^T\nabla^2 f \big|_{\hat{\theta}_n}(\theta-\hat{\theta}_n),
$$
and thus
\begin{align*}
\Psi_n^{h(gh)} &\Rightarrow \int \frac{\int \phi(\hat{\theta}_n \mid \theta , n^{-1}\hat{\Sigma}_n )h(\eta)d\eta  } {\int \phi(\hat{\theta}_n \mid \theta^0, n^{-1}\hat{\Sigma}_n )h(\eta) d\eta } g(\beta) d\beta\\
&=  \int \frac{ \phi(\hat{\beta}_n \mid \beta , n^{-1}\hat{\Sigma}_n^{\beta \beta} )\int \phi(\hat{\eta}_n \mid \eta + \hat{\Sigma}_n^{\eta \beta}(\hat{\Sigma}_n^{\beta\beta})^{-1} (\hat{\beta}_n-\beta), V_n) h(\eta)d\eta  } { \phi(\hat{\beta}_n \mid \beta^0 , n^{-1}\hat{\Sigma}_n^{\beta \beta} )\int \phi(\hat{\eta}_n \mid \eta + \hat{\Sigma}_n^{\eta \beta}(\hat{\Sigma}_n^{\beta\beta})^{-1} (\hat{\beta}_n-\beta^0), V_n) h(\eta) d\eta } g(\beta) d\beta
\end{align*}
where $V_n = n^{-1}\hat{\Sigma}_n^{\eta \eta}  - n^{-2}\hat{\Sigma}_n^{\beta \eta}(\hat{\Sigma}_n^{\eta \eta})^{-1}\hat{\Sigma}_n^{\eta \beta}$. Using the fact that $h$ is uniform, the inner integrals cancel out
\begin{align*}
\Psi_n^{h(gh)} &\Rightarrow \int \frac{ \phi(\hat{\beta}_n \mid \beta , n^{-1}\hat{\Sigma}_n^{\beta \beta} )\int \phi(\hat{\eta}_n \mid \eta + \hat{\Sigma}_n^{\eta \beta}(\hat{\Sigma}_n^{\beta\beta})^{-1} (\hat{\beta}_n-\beta), V_n)d\eta  } { \phi(\hat{\beta}_n \mid \beta^0 , n^{-1}\hat{\Sigma}_n^{\beta \beta} )\int \phi(\hat{\eta}_n \mid \eta + \hat{\Sigma}_n^{\eta \beta}(\hat{\Sigma}_n^{\beta\beta})^{-1} (\hat{\beta}_n-\beta^0), V_n)  d\eta } g(\beta) d\beta \\
&= \int \frac{ \phi(\hat{\beta}_n \mid \beta , n^{-1}\hat{\Sigma}_n^{\beta \beta} )  } { \phi(\hat{\beta}_n \mid \beta^0 , n^{-1}\hat{\Sigma}_n^{\beta \beta} ) } g(\beta) d\beta
\end{align*}
This proves the first part of the proposition. For the second proposition, we have that
\begin{align*}
\Psi_n^{h(gh')} &\Rightarrow \int \frac{ \phi(\hat{\beta}_n \mid \beta , n^{-1}\hat{\Sigma}_n^{\beta \beta} )\int \phi(\hat{\eta}_n \mid \eta + \hat{\Sigma}_n^{\eta \beta}(\hat{\Sigma}_n^{\beta\beta})^{-1} (\hat{\beta}_n-\beta), V_n) h(\eta + \Sigma^{\eta^0 \beta^0}(\Sigma^{\beta^0\beta^0})^{-1} (\beta_0 -\beta))d\eta  } { \phi(\hat{\beta}_n \mid \beta^0 , n^{-1}\hat{\Sigma}_n^{\beta \beta} )\int \phi(\hat{\eta}_n \mid \eta + \hat{\Sigma}_n^{\eta \beta}(\hat{\Sigma}_n^{\beta\beta})^{-1} (\hat{\beta}_n-\beta^0), V_n) h(\eta) d\eta } g(\beta) d\beta
\end{align*}
%Under the null hypothesis $\hat{\beta}_n \rightarrow \beta^0$ and $\hat{\Sigma}_n \rightarrow \Sigma$, thus
%\begin{align*}
%\Psi_n^{h(gh)} &\Rightarrow \int \frac{ \phi(\hat{\beta}_n \mid \beta , n^{-1}\hat{\Sigma}_n^{\beta \beta} )\int \phi(\hat{\eta}_n \mid \eta + \Sigma^{\eta \beta}(\Sigma^{\beta\beta})^{-1} (\beta^0-\beta), V_n) h(\eta + \Sigma^{\eta \beta}(\Sigma^{\beta\beta})^{-1} (\beta_0 -\beta))d\eta  } { \phi(\hat{\beta}_n \mid \beta^0 , n^{-1}\hat{\Sigma}_n^{\beta \beta} )\int \phi(\hat{\eta}_n \mid \eta, V_n) h(\eta) d\eta } g(\beta) d\beta
%\end{align*}
Consider the inner integrals of the numerator
$$
\phi(\hat{\eta}_n \mid \eta + \hat{\Sigma}_n^{\eta \beta}(\hat{\Sigma}_n^{\beta\beta})^{-1} (\hat{\beta}_n-\beta), V_n) h(\eta + \Sigma^{\eta^0 \beta^0}(\Sigma^{\beta^0\beta^0})^{-1} (\beta_0 -\beta))
$$
and denominator
$$
\phi(\hat{\eta}_n \mid \eta + \hat{\Sigma}_n^{\eta \beta}(\hat{\Sigma}_n^{\beta\beta})^{-1} (\hat{\beta}_n-\beta^0), V_n) h(\eta)
$$
define $\bar{\eta}_1(\beta)$ to be the maximizer of the numerator and $\bar{\eta}_0(\beta)$ the maximizer of the denominator term. Let $\eta^*(\beta) = \hat{\eta}_n - \hat{\Sigma}_n^{\eta \beta}(\hat{\Sigma}_n^{\beta\beta})^{-1} (\hat{\beta}_n-\beta)$ to be the maximum likelihood solution to $\phi(\hat{\eta}_n \mid \eta + \hat{\Sigma}_n^{\eta \beta}(\hat{\Sigma}_n^{\beta\beta})^{-1} (\hat{\beta}_n-\beta), V_n)$. Applying the Laplace approximation to the numerator and denominator yields.
\begin{align*}
\int \frac{ \phi(\hat{\beta}_n \mid \beta , n^{-1}\hat{\Sigma}_n^{\beta \beta} )\phi(\hat{\eta}_n \mid \bar{\eta}_1(\beta) + \hat{\Sigma}_n^{\eta \beta}(\hat{\Sigma}_n^{\beta\beta})^{-1} (\hat{\beta}_n-\beta), V_n) h(\bar{\eta}_1(\beta) + \Sigma^{\eta^0 \beta^0}(\Sigma^{\beta^0\beta^0})^{-1} (\beta_0 -\beta))|\Omega^1|^{\frac{1}{2}}} { \phi(\hat{\beta}_n \mid \beta^0 , n^{-1}\hat{\Sigma}_n^{\beta \beta} ) \phi(\hat{\eta}_n \mid \bar{\eta}_0(\beta^0) + \hat{\Sigma}_n^{\eta \beta}(\hat{\Sigma}_n^{\beta\beta})^{-1} (\hat{\beta}_n-\beta^0), V_n) h(\bar{\eta}_0(\beta^0)) |\Omega^0|^{\frac{1}{2}} } g(\beta) d\beta,
\end{align*}
where the $\Omega^i$ are the inverted second derivative matrices at $\bar{\eta}^i$. Because $\sqrt{n}(\bar{\eta}(\beta) - \eta^*(\beta)) \Rightarrow 0$ \citep{ghosh2007} we may substitute in $\eta^*$ resulting in
\begin{align*}
\Psi_n^{h(gh')} &\Rightarrow \int \frac{ \phi(\hat{\beta}_n \mid \beta , n^{-1}\hat{\Sigma}_n^{\beta \beta} )\phi(\hat{\eta}_n \mid \hat{\eta}_n, V_n) h(\hat{\eta}_n - \hat{\Sigma}_n^{\eta \beta}(\hat{\Sigma}_n^{\beta\beta})^{-1} (\hat{\beta}_n-\beta)+ \Sigma^{\eta^0 \beta^0}(\Sigma^{\beta^0\beta^0})^{-1} (\beta_0 -\beta))|\Omega^1|^{\frac{1}{2}}} { \phi(\hat{\beta}_n \mid \beta^0 , n^{-1}\hat{\Sigma}_n^{\beta \beta} ) \phi(\hat{\eta}_n \mid \hat{\eta}_n, V_n) h(\hat{\eta}_n - \hat{\Sigma}_n^{\eta \beta}(\hat{\Sigma}_n^{\beta\beta})^{-1} (\hat{\beta}_n-\beta^0)) |\Omega^0|^{\frac{1}{2}}}g(\beta) d\beta \\
&=\int \frac{ \phi(\hat{\beta}_n \mid \beta , n^{-1}\hat{\Sigma}_n^{\beta \beta} ) h(\hat{\eta}_n - \hat{\Sigma}_n^{\eta \beta}(\hat{\Sigma}_n^{\beta\beta})^{-1} (\hat{\beta}_n-\beta)+ \Sigma^{\eta^0 \beta^0}(\Sigma^{\beta^0\beta^0})^{-1} (\beta_0 -\beta))|\Omega^1|^{\frac{1}{2}}} { \phi(\hat{\beta}_n \mid \beta^0 , n^{-1}\hat{\Sigma}_n^{\beta \beta} ) h(\hat{\eta}_n - \hat{\Sigma}_n^{\eta \beta}(\hat{\Sigma}_n^{\beta\beta})^{-1} (\hat{\beta}_n-\beta^0)) |\Omega^0|^{\frac{1}{2}}}g(\beta) d\beta.
\end{align*}
Under the null hypothesis of  $\theta^0=\begin{bmatrix} \beta^0 \\ \eta^0 \end{bmatrix}$, $\hat{\beta}_n \rightarrow \beta^0$ and $\hat{\Sigma} \rightarrow \Sigma(\theta^0)$ so we may substitute these in within $h$.
$$
\Psi_n^{h(gh')} \Rightarrow \int \frac{ \phi(\hat{\beta}_n \mid \beta , n^{-1}\hat{\Sigma}_n^{\beta \beta} ) h(\hat{\eta}_n)|\Omega^1|^{\frac{1}{2}}} { \phi(\hat{\beta}_n \mid \beta^0 , n^{-1}\hat{\Sigma}_n^{\beta \beta} ) h(\hat{\eta}_n ) |\Omega^0|^{\frac{1}{2}}}g(\beta) d\beta.
$$
Similarly, under the null hypothesis of $\theta^0$, $\Omega^1 \rightarrow \Omega^0$. This can be seen by noting that the quantities inside the inner integrals of the numerator and denominator are asymptotically equivalent at $\eta^*(\beta)$ and $\eta^*(\beta^0)$ respectively. Thus we then have the result that
$$
\Psi_n^{h(gh')} \Rightarrow \int \frac{ \phi(\hat{\beta}_n \mid \beta , n^{-1}\hat{\Sigma}_n^{\beta \beta} ) } { \phi(\hat{\beta}_n \mid \beta^0 , n^{-1}\hat{\Sigma}_n^{\beta \beta} ) }g(\beta) d\beta.
$$

%As $n$ increases, the ratio is dominated by the normal likelihoods $\phi$, which become more and more concentrated, and so
%$$
%\Psi_n^{h(gh)} \Rightarrow  \int \frac{ \phi(\hat{\beta}_n \mid \beta , n^{-1}\hat{\Sigma}_n^{\beta \beta} )} { \phi(\hat{\beta}_n \mid \beta^0 , n^{-1}\hat{\Sigma}_n^{\beta \beta} ) }g(\beta) d\beta.
%$$

$\blacksquare$

\begin{theorem} \label{the:gmm}
Let $\hat{\theta}_{nk}$ be a (sub-)set of generalized method of moments estimators with limiting distribution $\sqrt{nk}\hat{\Sigma}_{nk}^{-\frac{1}{2}} (\hat{\theta}_{nk} - \theta) \Rightarrow N(0,I)$. The likelihood ratio 
$$
\Lambda_n = \frac{ \int \phi(\hat{\theta}_{nk} \mid \theta, (nk)^{-1}\hat{\Sigma})  g(\theta)d\theta} { \phi(\hat{\theta}_{nk} \mid  \theta_0, (nk)^{-1}\hat{\Sigma}) },
$$
is an asymptotically equivalent mSPRT as $n$ and $k>0$ become large. 
\end{theorem}
\textbf{Proof:}

The sample moment conditions for $\hat{\theta}_{nk}$ are
$$
m_n(\theta) = \frac{1}{n} \sum_i^n \gamma(X_i,\theta),
$$
and standard GMM asymptotic theory finds that
$$
\sqrt{nk} \Gamma^{-1}(\hat{\theta}_{nk} - \theta) \Rightarrow \sqrt{nk} m_{nk}(\theta),
$$
where $\Gamma$ is a constant matrix, and $m_{nk}(\theta) \Rightarrow N(0,nk\Omega)$ by the central limit theorem. Consider observing the sequence
$$
S_n = \Gamma \sum_{i=(n-1)k+1}^{nk}\gamma(X_i,\theta) + k\theta,
$$
then by construction 
$$
\sqrt{nk}\hat{\theta}_{nk} \Rightarrow \frac{\sum_i^nS_n}{\sqrt{nk}}.
$$

By the central limit theorem, for large $k$,
$$
S_n \Rightarrow N(k\theta,k\Gamma\Omega\Gamma^T).
$$
The sample mean and sample covariance are the maximum likelihood estimates of $k\theta$ and $k\Gamma\Omega\Gamma^T$, so we may apply Theorem \ref{the:asym} to obtain an asymptotically valid mSPRT. Because the sample mean and sample covariance are independent, we may then apply Theorem \ref{the:comp} setting $\beta=k\theta$ and $\eta=k\Gamma\Omega\Gamma^T$. This yields an mSPRT of
$$
\Lambda_n =  \frac{ \int \phi(\sum S_i \mid nk\theta, nk\hat{\Sigma})  g(\theta)d\theta} { \phi(\sum S_i \mid  nk\theta_0, nk\hat{\Sigma}) }.
$$ 
Replacing the $\sum S_i$ by the GMM estimator we arrive at
 $$
\Lambda_n =  \frac{ \int \phi(\hat{\theta}_{nk} \mid \theta, (nk)^{-1}\hat{\Sigma})  g(\theta)d\theta} { \phi(\hat{\theta}_{nk} \mid  \theta_0, (nk)^{-1}\hat{\Sigma}) }.
$$ 

$\blacksquare$

By Theorem \ref{the:gmm} we can still construct valid mSPRT tests for a wide variety of estimators by restricting our interim-analyses to occur every $k$ observations, provided $k$ is large enough. The result relies on the convergence of the central limit theorem within each batch of size $k$. In the case of online experiments, where tens of thousands of observations stream into a test per day, this it is not an unreasonable limitation. While a fixed $k$ interval is posited, the size of the interval may be allowed to grow or shrink over the course of the sequential test provided that each interval is long enough for the central limit theorem to apply to the batch.

\begin{corollary} \label{the:mlegmm}
Let $\hat{\theta}_{nk}$ be a (sub-)set of maximum likelihood estimators with limiting distribution $\sqrt{nk}\hat{\Sigma}_{nk}^{-\frac{1}{2}} (\hat{\theta}_{nk} - \theta) \Rightarrow N(0,I)$. The likelihood ratio 
$$
\Lambda_n = \frac{ \int \phi(\hat{\theta}_{nk} \mid \theta, (nk)^{-1}\hat{\Sigma})  g(\theta)d\theta} { \phi(\hat{\theta}_{nk} \mid  \theta_0, (nk)^{-1}\hat{\Sigma}) },
$$
is an asymptotically equivalent mSPRT as $n$ and $k>0$ become large. 
\end{corollary}

Corollary \ref{the:mlegmm} follows immediately from the fact that maximum likelihood estimators are also generalized method of moments estimators. Thus we now have three ways to justify the use of the asymptotic marginal distribution of a subset of the MLEs. Firstly, if the subset of estimators is independent of the rest of the estimators, then Theorem \ref{the:comp} may be used to remove the nuisance parameters. Otherwise, Theorem \ref{the:comp2} shows that mSPRT limits to $L_n^{hg}$, which controls the averaged type I error of the test. Finally, Corollary \ref{the:mlegmm} shows that (non-averaged) type I error is controlled so long as the mSPRT is evaluated at intervals. The fact that the mSPRT is valid when evaluated at intervals suggests that type I errors are likely to be well controlled even when $k=1$, as the deviation of the test statistic within each interval is bounded.

\begin{theorem} \label{the:anc}
Let $Y_1,...,Y_\infty$ be an ancillary sequence of  random variables such that 
$$
p(X_1,...,X_n,Y_1,...,Y_n) = \prod_{i=1}^np(X_i \mid \theta, Y_i)p(Y_i | Y_1,...,Y_{i-1}).
$$
then
$$
\Lambda_n^g = \frac{ \int  p(X_1,...,X_n \mid \theta, Y_1,...,Y_n)  g(\theta)d\theta} { p(X_1,...,X_n \mid \theta_0, Y_1,...,Y_n) },
$$
\end{theorem}
\textbf{Proof:}
\begin{align*}
\Lambda_n^g &= \frac{ \int \prod_{i=1}^np(X_i \mid \theta, Y_i)p(Y_i | Y_1,...,Y_{i-1})  g(\theta)d\theta} { \prod_{i=1}^np(X_i \mid \theta_0, Y_i)p(Y_i | Y_1,...,Y_{i-1}) }\\
&= \frac{ \int \prod_{i=1}^np(X_i \mid \theta, Y_i)  g(\theta)d\theta} { \prod_{i=1}^np(X_i \mid \theta_0, Y_i) } \\
&= \frac{ \int  p(X_1,...,X_n \mid \theta, Y_1,...,Y_n)  g(\theta)d\theta} { p(X_1,...,X_n \mid \theta_0, Y_1,...,Y_n) }.
\end{align*}

$\blacksquare$

\section{An Efficient Family of Priors}

Thus far we have avoided putting a defined functional form on the prior under the alternative hypothesis $g$. While in principle any distribution may be selected, computing the mixture integral numerically for each observation can be prohibitively computationally expensive, especially in the case of online experiments, where the number of observations is typically greater than 10,000, and may scale up to the millions. Fortunately, we may specify a family of distributions that is both flexible enough to approximate any arbitrary distribution while providing a closed form solution to the integral.

Let $g'$ be a multivariate mixture normal density with $r$ components
$$
g'(\theta) = \sum_{i=1}^r \phi(\theta | \mu_i, \Upsilon_i)w_i,
$$
 where $w_i$ is the probability of selecting the $i$th component. Let $\Lambda_n$ be a mSPRT with normal likelihoods of the form
 $$
\frac{ \int \phi(\hat{\theta}_{n} \mid \theta, \hat{\Sigma}_n)  g(\theta)d\theta} { \phi(\hat{\theta}_{k} \mid  \theta_0, \hat{\Sigma}_n) }.
$$ 
Given $g=g'$, the mSPRT simplifies to
$$
\frac{ \sum_i^r \phi(\hat{\theta}_{n} \mid \mu_i, \hat{\Sigma}_n + \Upsilon_i)  w_i} { \phi(\hat{\theta}_{k} \mid  \theta_0, \hat{\Sigma}_n) },
$$
removing the need for numeric integration. The mixture normal distribution allows us to efficiently represent most distributions with just a few components and also allows us the flexibility to model any continuous distribution by simply increasing the number of terms.

\section{Applications in Online Testing}

\subsection{$m$-Sample Tests of a Binomial Outcome}

One of the most common goals for an online A/B test is to determine whether one arm of the trial leads to more ``conversions'' then the others. A conversion might indicate signing up for a news letter, a purchase, clicking on an Ad, or any other positive action by the user. The outcome is therefore a Bernoulli random variable $X_i \sim \textrm{Ber}(p_{Y_i})$, where $Y_i \in \{1,...,m\}$ is the arm assigned to the $i$th individual. The maximum likelihood estimators of $p$ are simply the sample proportions within each group $\hat{p}_j = \frac{1}{n_j}\sum_iX_i\mathds{1}(Y_i=j)$, where $\mathds{1}$ is the indicator function and $n_j=\sum_i \mathds{1}( Y_i=j)$.

\subsubsection{Risk Ratio}
Since $\log$ is a one-to-one function, the estimators $\log(\hat{p})$ are the maximum likelihood estimators of $\log(p)$ and by the delta method are independent asymptotically normal variables with variance 
$$
v_j=\hat{\textrm{var}}(\log(\hat{p}_j))=\frac{1-\hat{p}_j}{\hat{p}_jn_j}.
$$

Defining $\beta^i = \log(p_{i+1}) - \log(p_1)$ to be the log risk ratio versus baseline, $\eta=\sum_j \log(p_j)$ to be a nuisance parameter, $\hat{\theta}_n=\begin{bmatrix} \hat{\eta}_n \\ \hat{\beta}_n \end{bmatrix}$ to be the maximum likelihood estimates of the re-parameterized model obtained by replacing $p$ by $\hat{p}$. The covariance between $\beta^i$ and $\eta$ is $v_i - v_1$. This covariance tends to 0 under the the null hypothesis of $H_0:\beta=0$ if the allocation rates are balanced across all variants. Therefore a consistent estimate of the covariance under $H_0$ and balanced allocation is
$$
n^{-1}\hat{\Sigma}_n = \begin{bmatrix} \sum_j v_j & 0 \\ 0 & n^{-1}\hat{\Sigma}^{\beta \beta}_n \end{bmatrix},
$$
where $n^{-1}(\hat{\Sigma}^{\beta \beta}_n)_{jj} = v_{j+1}+v_1$ and $n^{-1}(\hat{\Sigma}^{\beta \beta}_n)_{ij} = v_1 \ \ \forall i \neq j$. We may now apply Theorems \ref{the:asym} and \ref{the:comp} to obtain an asymptotically valid mSPRT
$$
\frac{ \int \phi(\hat{\beta}_n \mid \beta , n^{-1}\hat{\Sigma}_n^{\beta \beta} )  g(\beta) d\beta} { \phi(\hat{\beta}_n \mid 0, n^{-1}\hat{\Sigma}_n^{\beta \beta} ) }.
$$

Because $\hat{\beta}^i_n$ are maximum likelihood estimators for the reparametrized model, when allocations are unbalanced or adaptive allocation, Theorem \ref{the:comp2} and Corollary \ref{the:mlegmm} may be used to justify the mSPRT.

\subsubsection{Odds Ratio}

Derivation of an mSPRT for the log odds ratio proceeds similarly to the risk ratio case. Setting $\beta^i = \log(\frac{p_{i+1}}{1-p_{i+1}}) - \log(\frac{p_{1}}{1-p_{1}})$ to be the log odds ratio versus baseline, $\eta=\sum_j \log(\frac{p_j}{1-p_j})$ to be a nuisance parameter we apply Theorems \ref{the:asym} and \ref{the:comp} to obtain an asymptotically valid mSPRT in the case of balanced allocations and utilize Theorem \ref{the:comp} and Corollary \ref{the:mlegmm} to justify it in the case of unbalanced or adaptive allocations.
$$
\frac{ \int \phi(\hat{\beta}_n \mid \beta , n^{-1}\hat{\Sigma}_n^{\beta \beta} )  g(\beta) d\beta} { \phi(\hat{\beta}_n \mid 0, n^{-1}\hat{\Sigma}_n^{\beta \beta} ) }.
$$
where $n^{-1}(\hat{\Sigma}^{\beta \beta}_n)_{jj} = v_{j+1}+v_1$ and $n^{-1}(\hat{\Sigma}^{\beta \beta}_n)_{ij} = v_1 \ \ \forall i \neq j$ for $v_i=\frac{1}{n_i\hat{p}_i} + \frac{1}{n_i(1-\hat{p}_i)}$.

\subsubsection{Difference in Proportions}

We begin our mSPRT for difference in proportions similarly to the risk and odds ratio cases, by defining $\beta^i=p_{i+1}-p_1$ and $\eta=\sum_i p_i$. By an identical argument,
$$
\frac{ \int \phi(\hat{\beta}_n \mid \beta , n^{-1}\hat{\Sigma}_n^{\beta \beta} )  g(\beta) d\beta} { \phi(\hat{\beta}_n \mid 0, n^{-1}\hat{\Sigma}_n^{\beta \beta} ) }
$$
is an asymptotically valid mSPRT under balanced allocation, where $n^{-1}(\hat{\Sigma}^{\beta \beta}_n)_{jj} = v_{j+1}+v_1$ and $n^{-1}(\hat{\Sigma}^{\beta \beta}_n)_{ij} = v_1 \ \ \forall i \neq j$ for $v_i=\frac{\hat{p}_i(1-\hat{p}_i)}{n_i}$. Again, we utilize Theorem \ref{the:comp2} and Corollary \ref{the:mlegmm} to justify it in the case of unbalanced or adaptive allocations.

When there are only two arms to the trial, this reduces to the difference in proportions test described in \cite{opt1}; However, the derivation of the test in that paper is incorrect. Lemma 4 of \cite{opt1} first asserts that $\hat{p}_2-\hat{p}_1$ is asymptotically independent of $\hat{p}_2+\hat{p}_1$ for any $p$, whereas this is only true under the null hypothesis of equality of proportions and equal allocation rates since $\textrm{cov}(\hat{p}_2-\hat{p}_1, \hat{p}_2+\hat{p}_1) = v_2 - v_1$. Secondly, they assert that $\hat{\Sigma}^{\beta \beta}_n$ converges to both $\Sigma^{\beta \beta}(\beta=0)$ and $\Sigma^{\beta \beta}(\beta) \ \ \forall \beta \neq 0$ in the same experiment, which of course it can not do because nothing can converge to two different matrices.

Another problem with difference in proportion is that it is a poor measure of the effect of an intervention. For example, increasing the conversion rate of a page from 50\% to 51\% is a small change, likely to lead to a small increase in the underlying profit. A change from a 1\% conversion rate to a 2\% conversion rate on the other hand is a huge effect, which (if revenue is linked to conversions) could double profit. For this reason, difference in proportion is generally of less business use than risk ratio.

Because difference in proportion effect size varies based on the base rate, identifying a $g$ that is useful for both high and low conversion experiments is a challenge. Our approach is to model the prior of the Cohen's D effect size \citep{hedges2014} instead of the raw proportion. Effect size is defined as 
$$
\frac{p_j-p_1}{\bar{\sigma}},
$$
where $\bar{\sigma}$ is a measure of the underlying variation, and could be the baseline standard deviation $\sqrt{p_1(1-p_1)}$, the average variation across groups $\sqrt{\frac{1}{m}\sum_i p_i(1-p_i)}$, or any other population quantity that can be estimated consistently. Then 
$$
\frac{ \int \phi(\hat{\beta}_n \mid \beta , n^{-1}\hat{\Sigma}_n^{\beta \beta} )  g(\frac{\beta}{\hat{\sigma}_n}) d\beta} { \phi(\hat{\beta}_n \mid 0, n^{-1}\hat{\Sigma}_n^{\beta \beta} ) } \Rightarrow \frac{ \int \phi(\hat{\beta}_n \mid \beta , n^{-1}\hat{\Sigma}_n^{\beta \beta} )  g(\frac{\beta}{\bar{\sigma}}) d\beta} { \phi(\hat{\beta}_n \mid 0, n^{-1}\hat{\Sigma}_n^{\beta \beta} ) },
$$
where $\hat{\sigma}_n$ is a consistent approximation of $\bar{\sigma}$ for example $\sqrt{\hat{p}_1(1-\hat{p}_1)}$ or $\sqrt{\frac{1}{m}\sum_i \hat{p}_i(1-\hat{p}_i)}$.

\subsection{$m$-Sample Tests of a Numeric Outcome}

For numeric outcomes, instead of being distributed binomially, the outcome is distributed according to $X_i \sim f_{Y_i}$, where $f_{Y_i}$ is the distribution of the $i$th arm of the study. Of particular interest is the case when each group is normally distributed $f_{Y_i}(X_i) = \phi(X_i | \mu_{Y_i}, \sigma^2_{Y_i})$.

\subsubsection{Difference in Means}

Let us begin by considering the simplified case of $f_{Y_i}(X_i) = \phi(X_i | \mu_{Y_i}, \sigma^2_{Y_i})$. The maximum likelihood estimates are the group means and standard deviations $\hat{\mu}_j = \frac{1}{n_j}\sum_iX_i\mathds{1}(Y_i=j)$ and $\hat{\sigma}^2_j = \frac{1}{n_j-1} \sum_i(X_i - \hat{\mu}_j)\mathds{1}(Y_i=j)$. We begin by reparameterizing the model as $\beta^i = \mu_{i+1} - \mu_1$, $\eta = (\sum_i \mu_i, \sigma^2_1,...,\sigma^2_n)$.

The MLEs for $\mu$ and $\sigma^2$ are the sample means and variances, which under the normal model are independent. Under the null hypothesis of equality of means, if the allocations are balance and $\sigma^2_i$ are all equal then the asymptotic covariance between $\hat{\beta}_n$ and $\hat{\eta}_n$ are all zero. Therefore we may apply Theorems \ref{the:asym} and \ref{the:comp} to obtain 
$$
 \frac{ \int \phi(\hat{\beta}_n \mid \beta , n^{-1}\hat{\Sigma}_n^{\beta \beta} )  g(\beta) d\beta} { \phi(\hat{\beta}_n \mid 0, n^{-1}\hat{\Sigma}_n^{\beta \beta} ) },
$$
where $n^{-1}(\hat{\Sigma}^{\beta \beta}_n)_{jj} = v_{j+1}+v_1$ and $n^{-1}(\hat{\Sigma}^{\beta \beta}_n)_{ij} = v_1 \ \ \forall i \neq j$ for $v_i=\frac{\hat{\sigma}^2_i}{n_i}$.

Because $\hat{\beta}^i_n$ are maximum likelihood estimators for the reparametrized normal model, when allocations are unbalanced and/or the true variances are unequal, Theorem \ref{the:comp2} and Corollary \ref{the:mlegmm} may be used to justify the mSPRT.

In reality, very few outcomes in online testing applications are even approximately normally distributed. Many have heavy tails and high skew. Fortunately, Theorem \ref{the:gmm} provides a foundation for inference in the case of non-normal data. Firstly note that from the central limit theorem we have that $\hat{\Sigma}^{\beta \beta}_n$ remains a consistent estimator of the asymptotic covariance. Let $Y_n$ be distributed multinomial with probabilities $q$, and consider
$$
\gamma(X_i,Y_i,\mu_j - \mu_1) = \frac{X_i\mathds{1}(Y_i=j)}{q_j} - \frac{X_i\mathds{1}(Y_i=1)}{q_1} - (\mu_j - \mu_1).
$$
The expectation of $\gamma$ is 0, so
$$
\frac{1}{n}\sum_i \gamma(X_i,Y_i,\mu_j - \mu_1) = 0
$$
is a (generalized) method of moments estimator with solution
$$
(\hat{\beta}^j_n)_{\textrm{gmm}} = \frac{1}{n}\sum_i \big(\frac{X_i\mathds{1}(Y_i=j)}{q_j} - \frac{X_i\mathds{1}(Y_i=1)}{q_1}\big) = \mu_j - \mu_1.
$$
Because $n_j \Rightarrow nq_j$, we have that $\sqrt{n}\hat{\beta}^j_n \Rightarrow \sqrt{n}(\hat{\beta}^j_n)_{\textrm{gmm}}$, and so the maximum likelihood estimators under the normal model are asymptotically equivalent to GMM estimators under the general distribution model. Thus, we may use Theorem \ref{the:gmm} as a basis for inference.

As with differences in proportions, the scale of $X$ is an important factor in determining the likely prior distribution effect sizes and again the solution is to model the prior effect size using Cohen's D type statistics 
$$
\frac{\mu_j-\mu_1}{\bar{\sigma}},
$$
where $\bar{\sigma}$ here equals the population quantity that may be estimated consistently; for example, the variance of the baseline variant ($\textrm{var}_{f_1}(X)$) or the average variance across variants ($\frac{1}{m}\sum_j^m\textrm{var}_{f_j}(X)$). Then 
$$
\frac{ \int \phi(\hat{\beta}_n \mid \beta , n^{-1}\hat{\Sigma}_n^{\beta \beta} )  g(\frac{\beta}{\hat{\sigma}_n}) d\beta} { \phi(\hat{\beta}_n \mid 0, n^{-1}\hat{\Sigma}_n^{\beta \beta} ) } \Rightarrow \frac{ \int \phi(\hat{\beta}_n \mid \beta , n^{-1}\hat{\Sigma}_n^{\beta \beta} )  g(\frac{\beta}{\bar{\sigma}}) d\beta} { \phi(\hat{\beta}_n \mid 0, n^{-1}\hat{\Sigma}_n^{\beta \beta} ) } ,
$$
where $\hat{\sigma}$ is the sample estimate of $\bar{\sigma}$, for example $\hat{\sigma}_n = \sqrt{\frac{1}{n_1-1} \sum_i(X_i-\hat{\mu}_n)^2\mathds{1}(Y_i=1)}$.

\subsection{Non-parametric Superiority}

As mentioned in the previous section. Many applications in online testing involve outcomes with heavy tails and high skew. The presence of outliers due either to exceptional users, or data collection errors may factor into choosing appropriate methodologies for analysis.

For heavy tailed distributions with outliers, it is well known that the mean, as a measure of central tendency, is a questionable choice. The mean is heavily influenced by the tail behavior of a distribution, and thus any test based on mean differences will require large sample sizes to reach significance. Further, the result of that test may be dominated by the behavior of a minority of exceptional users rather than representing the effects of the experiment on the majority.

Another important use case is ordinal data, which, while ordered, does not have an intrinsic unit of measurement. Examples from online testing might be the number of steps a user took through the registration process, or a user selected product rating from ``Very good'' to ``Very Poor.'' Using means to measure the central tendency of an ordinal variable imposes an arbitrary unit on the variable, which may not be appropriate.

Addressing both the continuous and ordinal case in the comparison of two samples is known as the Nonparametric Behrens-Fisher Problem \citep{brunner2000}. In the non-sequential context \cite{brunner2000} developed a two-sample test that shows good small sample characteristics.

Let $g_i$ be the indexes of $X$ belonging to group $i$, then the treatment effect of group $i$ over group $1$ is defined as
$$
p_i = P(X_{(g_1)_1} < X_{(g_i)_1}) + \frac{1}{2}P(X_{(g_1)_1} = X_{(g_i)_1}).
$$
The interpretation of this treatment effect is that $p_i$ is the probability that a random chosen member of group $i$ has a higher value of $X$ than a randomly chosen member of group 1, plus 0.5 times the probability that they tie. If the two distributions are equal, then $p_i=0.5$. If group $i$ tends to have higher values, then $0.5 < p_i \leq 1$ and if it tends to have lower values then $0 \leq p_i < 0.5$. $p_i$ is also known as the area under the curve (AUC) \citep{mason2002}.

Following \cite{brunner2000}, we define the normalized distribution function $F_i(x) = \frac{1}{2}\big(F_i^-(x) + F_i^+(x)\big)$, where $F_i^-(x) = P(X_{(g_i)_1} < x)$ is the left continuous distribution function and $F_i^+(x) = P(X_{(g_i)_1} \leq x)$ is the right continuous version. Empirical approximations $\hat{F}$ of $F$ may be estimated by replacing the probabilities by their sample analogs. Further, we define the mid-rank of each $X_j$ as $R_j$, and $\bar{R}_i = \frac{1}{n_i}\sum_{j \in g_i}R_j$ to be the observed mean rank of group $i$. An unbiased estimate of $p_i$ is then
$$
\hat{p}_i=\int\hat{F}_1d\hat{F}_i = \frac{1}{n_1} \Bigg(\bar{R}_i - \frac{n_i-1}{2}\Bigg).
$$

\cite{brunner2000} then make large sample inference possible by showing that asymptotically,
\begin{equation}\label{eq:bm}
\sqrt{n}(\hat{p}_i - p_i) \Rightarrow U_n = \sqrt{n}\Bigg(\frac{1}{n_i}\sum_{j\in g_i}F_1(X_j) - \frac{1}{n_1}\sum_{j\in g_1}F_i(X_j) + 1 - 2p_i \Bigg).
\end{equation}

The right hand side of Equation \ref{eq:bm} is the difference of two sums of independent variables, and thus the central limit theorem may be invoked for asymptotic normality. The variance may be expressed as
$$
\textrm{var}(U_n) = n\Bigg(\frac{v_i^2}{n_1} + \frac{\sigma_i^2}{n_i}\Bigg),
$$
where $\sigma_i^2 = \textrm{var}(F_i(X_{(g_1)_1}))$ and $v^2 = \textrm{var}(F_1(X_{(g_i)_1}))$. $\sigma$ and $v$ may be consistently approximated by $\hat{\sigma}_i^2 = \hat{\textrm{var}}(\hat{F}_i(X_{(g_1)_1}))$ and $\hat{v}^2 = \hat{\textrm{var}}(\hat{F}_1(X_{(g_i)_1}))$, where $\textrm{var}$ is the sample variance. So, the asymptotic distribution of $\hat{p}_i$ may be approximated as
$$
\hat{p}_i \sim N\Big(p_i, \frac{\hat{v}_i^2}{n_1} + \frac{\hat{\sigma}_i^2}{n_i}\Big)
$$

\subsubsection{Extending the Test to Sequential Data}

Unlike the discrete, or normal distribution cases, $\hat{p}$ are not maximum likelihood estimators. The asymptotic distribution of $\hat{p}$ is normal, and the diagonal terms of the covariance matrix are
$$
n^{-1}\hat{\Sigma}_{ii} = \frac{\hat{v}_i^2}{n_1} + \frac{\hat{\sigma}_i^2}{n_i}.
$$
The off-diagonal terms may be estimated noting that
\begin{align*}
    \textrm{cov}(\hat{p}_l,\hat{p}_m) &\Rightarrow \textrm{cov}\Bigg( \frac{1}{n_l}\sum_{j\in g_l}F_1(X_j) - \frac{1}{n_1}\sum_{j\in g_1}F_l(X_j), \frac{1}{n_m}\sum_{j\in g_m}F_1(X_j) - \frac{1}{n_1}\sum_{j\in g_1}F_m(X_j) \Bigg) \\
    &=\textrm{cov}\Bigg( \frac{1}{n_1}\sum_{j\in g_1}F_l(X_j),\frac{1}{n_1}\sum_{j\in g_1}F_m(X_j) \Bigg) \\
    &= \frac{1}{n_1}\textrm{cov}\Bigg(F_l(X_{(g_1)_1}),F_m(X_{(g_1)_1}) \Bigg) \\
    &\approx n^{-1}\hat{\Sigma}_{lm}= \frac{1}{n_1}\hat{\textrm{cov}}\Bigg(\hat{F}_l(X_{(g_1)_1}),\hat{F}_m(X_{(g_1)_1}) \Bigg),
\end{align*}
where $\hat{\textrm{cov}}$ is the sample covariance. In order for the asymptotic mSPRT
\begin{equation}\label{eq:sup}
\frac{ \int \phi(\hat{p} \mid p , n^{-1}\hat{\Sigma} )  g(p) dp} { \phi(\hat{p} \mid \frac{1}{2}, n^{-1}\hat{\Sigma} ) },
\end{equation}
to be justified, we must show that $\hat{p}$ is asymptotically equivalent to a generalized method of moments estimator. As with the derivation with regard to difference in means, we let $Y_n$ be distributed multinomial with probabilities $q$. Consider the function
$$
\gamma_j(X_i,Y_i,p_j) = \frac{F_1(X_i)\mathds{1}(Y_i=j)}{q_j} - \frac{F_j(X_i)\mathds{1}(Y_i=1)}{q_1} + 1 - 2p_j.
$$
From \cite{brunner2000} we know that $E(\gamma_j(X_i,Y_i,p_j))=0$, and thus the solution to
$$
\frac{1}{n}\sum_i \gamma_j(X_i,Y_i,p_j) = 0
$$
is a generalized method of moments estimator. Because $\frac{n_i}{n} \Rightarrow q_i$,
\begin{align*}
    \sqrt{n}(\hat{p}_j - p_j) &\Rightarrow U_n \\
    &\Rightarrow \sqrt{n} \frac{1}{n}\sum_i \gamma_j(X_i,Y_i,p_j),
\end{align*}
and thus the solution $\hat{p}_j$ is asymptotically equivalent to a generalized method of moments estimator. Theorem \ref{the:gmm} may then be applied to justify Equation \ref{eq:sup} as a sequential mSPRT.

\section{Discussion}

The field of A/B testing is just recently realizing the importance of sequentially valid inference. Utilizing traditional statistical equations in an environment where the analyst is checking for significance daily leads to wildly inflated type I errors. It is thus important to understand how sequentially valid inference may be applied to the performance indicators and metrics of interest to the community.

The methodology developed here allows for any performance indicator that can be seen as either a maximum likelihood estimator, or a generalized method of moments estimator to be formulated into a sequentially valid test provided a estimate of it's standard error is available.

\bibliographystyle{plainnat}
\bibliography{sequential}

\end{document}